\documentstyle[pre,aps,epsfig,multicol]{revtex}
\begin{document}
\title{Colloidal dipolar interactions in 2D smectic C films}

\author{Pedro Patr\'\i cio, M. Tasinkevych
and M.M. Telo da Gama} 

\address{Departamento de F{\'\i}sica da Faculdade de Ci{\^e}ncias and 
Centro de F{\'\i}sica Te\'orica e Computacional\\
Universidade de Lisboa, 
Avenida Professor Gama Pinto 2, P-1649-003 Lisboa Codex, Portugal}

\date{July 9, 2001}
 
\maketitle
 
\begin{abstract}

We use a two-dimensional (2D) elastic free energy to calculate the
effective interaction between two circular disks immersed in smectic-$C$
films.
For strong homeotropic anchoring, the distortion of
the director field caused by the disks generates additional
topological defects that induce an effective interaction between the
disks. We use finite elements, with adaptive meshing, to minimize the 2D
elastic free energy. The method is shown to be accurate and efficient for
inhomogeneities on the length scales set by the disks and the defects,
that differ by up to 3 orders of magnitude.
We compute the effective interaction between two disk-defect pairs 
in a simple (linear) configuration. 
For large disk separations, $D$, the elastic free energy scales as $\sim
D^{-2}$, confirming the dipolar character of the long-range effective
interaction. For small $D$ the energy exhibits a pronounced minimum. The
lowest energy corresponds to a symmetrical configuration of the
disk-deffect pairs, with the inner defect at the mid-point between the
disks. The disks are separated by a distance that is twice the distance
of the outer defect from the nearest disk. The latter is identical to the
equilibrium distance of a defect nucleated by an isolated disk. 
\end{abstract}

\begin{multicols}{2}

\section{Introduction}
 
Colloidal systems consist of a dispersive medium or solvent and a
dispersed phase. The size of the dispersed domains or colloids is
large compared to atoms but small compared to macroscopic lengths.
Thus these systems exhibit new phenomena that are challenging
theoretically while they are of practical importance in a variety of 
applications, from paints and coatings to food and drugs \cite{Russel}. 

The colloidal state is metastable and its life-time is determined
(largely) by the effective interactions between colloids. In many systems
these interactions are tunable through the control of one or more external
parameters. 
To prevent phase separation due to van der Waals or other attractive
interactions, the colloids are coated or charged to induce steric
or electrostatic repulsions \cite{Russel}. Recently,
a novel mechanism of colloidal interactions was reported: the elastic
distortion of a liquid crystal host in a reverse nematic emulsion
\cite{poulin_science},
i.e., an emulsion of water droplets in a nematic liquid. 
This interaction arises from a competition between the
aligning properties of the spherical droplet surfaces that favour
e.g., a radial (homeotropic) nematic orientation and
the bulk elasticity and boundary conditions that favour a uniform
nematic. This effect is generic and it arises in a variety of
anisotropic liquids for different isotropic inclusions. In addition, 
elastic mediated interactions were shown to be effective above the bulk
nematic-isotropic temperature, for colloids wet by the nematic phase
\cite{Galatola2001}. 
Finally, under appropriate conditions, inverted nematic emulsions exhibit
colloidal self-organization, i.e., new phases where the colloids 
are ordered over macroscopic distances \cite{loudet,loudet_epl}. 

Understanding the phase behaviour of these complex systems begins by
identifying the mechanisms and calculating the effective
interactions between colloids. This is far from a simple task
owing to the non-linear nature of the nematic elasticity and the existence
of widely different length scales: the bulk correlation length that sets
the
scale for variations of the microscopic degrees of freedom, such as
nematic order, biaxility etc., and the size of the colloids
that sets the scale for variations of the elastic degrees of freedom,
i.e., the nematic director field, $\bf n(r)$  \cite{Chaikin}.
Both microscopic (molecular dynamics simulations \cite{andrienko}, 
tensor order-parameter free energy \cite{fukuda}) and
macroscopic (elastic free energy minimization using: trial
functions \cite{poulin_science,lubensky}, simulated anealing
\cite{ruhwandl}, finite elements \cite{stark} and electrostatic analogies
\cite{pettey}) approaches have been used to calculate the orientational
order of a nematic in the presence of a single colloid, in 2 and 3 dimensions. 

Theoretical \cite{lubensky,ruhwandl,stark} and experimental
\cite{poulin_science,gu,mondain-monval} work on inverted nematic emulsions
in 3D, reveals that for sufficiently strong homeotropic anchoring, spherical 
droplets induce topological defects, i.e., singularities in the
nematic director field.
There are several of these defects and their stability depends on 
parameters such as the size of the droplet, the anchoring strength and the
boundary conditions. 
When the size of the droplet is large, a hyperbolic point defect appears
at a certain distance from the droplet, and this structure is
known as the {\sl satellite} configuration \cite{lubensky,ruhwandl,stark}.
For smaller droplets and/or weaker anchoring strengths, the minimum 
elastic free energy corresponds to a ring disclination that surrounds
the droplet on the equatorial plane, and this structure is known as the
{\sl saturn-ring} configuration \cite{lubensky,ruhwandl,stark}.  
The distortion of the director field due to the presence of more
than one of these colloids induces multipolar long-range interactions
between the colloids \cite{poulin_science,poulin,stark2}.
Depending on the symmetry of the distortion, the dominant long-range
interactions are dipolar \cite{poulin} or quadrupolar 
\cite{kuksenok,mondain-monval}.
At short range, however, the colloids repel each other
\cite{stark2} and this prevents their colapse.

A 2D system was also investigated by Pettey et al. \cite{pettey} who
obtained analytic solutions for the smectic-$C$ director around circular
disks.
For a disk at the center of a smectic-$C$ layer, the solution that
minimizes
the elastic free energy exhibits a hyperbolic point defect, at a distance 
$r_d^0=\sqrt{2}a$ from the center. 
The free energy of a system of multiple disk-defect pairs was also
obtained, for pairs that are far apart. For defects aligned
with the far field director, a dipole-dipole interaction results between
the colloids \cite{pettey}.
These predictions were tested in a recent experiment by Cluzeau et
al. \cite{cluzeau}. Thin films of smectic-$C^*$ were
heated and the temperature was stabilized when cholesteric 
($N^*$) droplets started to nucleate.
Cluzeau and collaborators observed that a hyperbolic point defect
is nucleated with each cholesteric droplet, at the distance predicted
by theory.
In addition, they observed that the droplets interact through a long-range
attraction and assemble into chains, confirming the dipolar nature of the
interaction.
Finally, they measured a mean separation $D=(1.3\pm 0.1)d_a$ between
nearest-neighbor droplets in chains, where $d_a$ is the droplet average
diameter.

Even in 2D exact solutions are not readily obtained. The analytic 
solutions of \cite{pettey} are restricted to one inclusion or to multiple
inclusions that are sufficiently far apart.    
Recently, Fukuda and Yokoyama solved the 2D problem using the tensor order
parameter formalism and an efficient adaptive grid method to integrate
numerically the dynamic equation \cite{fukuda}. 
One of the advantages of this method is that no special treatment of 
topological defects is required, by contrast with the director
description, where defects appear as singularities of the director field 
\cite{Chaikin}. 
For all the parameters used in \cite{fukuda}, the authors found that the
equilibrium director configuration around a circular disk, exhibits 
quadrupolar symmetry and corresponds to two defects with topological
charge -1/2. This contrasts 
with the 3D result, where the quadrupolar defect was found to be stable over 
a limited range of parameters. It is also in apparent contradiction with
the experiment on 2D smectic-$C^*$ films \cite{cluzeau}. The latter,
however, do not exhibit mirror symmetry about reflection of the
2D director, excluding a distortion with quadrupolar symmetry 
\cite{degennes,Chaikin}. 

In this paper we use finite elements and adaptive meshes to minimize the
2D elastic free energy functional. We focus on dipolar distortion
fields as observed in the experiments \cite{cluzeau}. 
In particular, we study the interaction between two disks of equal radii,
$a$, in a large smectic-$C$ layer.
The minimal free energy occurs when the two dipoles are aligned head to
tail, in a direction parallel to the far field director. In this geometry
we investigated a wide range of disk separations. We verified the validity
of the electrostatic analogy at large distances and calculated
the short-range repulsive interaction between two disks.

\section{Model and numerical method}

The free energy of a nematic is invariant under uniform 
rotations of the whole sample and under the symmetry operations $\bf n
\rightarrow  -n$ and $\bf r \rightarrow -r$. In addition, $\bf n$ is a unit 
vector so that $n_i \nabla_j n_i$ is zero. These considerations imply that the 
stiffness tensor has 2 independent bulk components in 2D. The distortions
whose energy is measured by these constants are (1) splay, with non-zero 
$\nabla \cdot \bf n$, and (2) bend, with non-zero 
$\nabla \times \bf n$. The elastic free energy for the bulk 
nematic phase, is the sum of these terms \cite{Frank,Chaikin}: 
\begin{equation}
F=\frac{1}{2} \int [ K_S (\nabla \cdot {\bf n})^2 + 
K_B ( \nabla \times {\bf n})^2 ] dS
\label{eq:Frank}
\end{equation}
where $K_S$ and $K_B$ are the elastic constants associated
with splay and bend distortions, and 
${\bf n}=(\cos\theta,\sin\theta)$
is the 2D nematic director.
The integral is over the region occupied by the 2D nematic film.

Note that if $K_S=K_B=K$, the elastic free energy is
proportional to $\sum_{ij} \nabla_i n_j \nabla_i n_j$ (plus boundary
terms) which is the form produced by simple models of the nematic state, such 
as those with pairwise intermolecular interactions depending only on the 
angle between the long molecular axes or the 2D XY-model \cite{pettey}.
\begin{eqnarray}
F&=&\frac{K}{2} \int [ (\nabla \cdot {\bf n})^2 + 
( \nabla \times {\bf n})^2 ] dS\\
&=&\frac{K}{2} \int (\nabla\theta)^2dS
\label{eq:model}
\end{eqnarray}
Line and boundary terms have been ignored in (\ref{eq:model}) which is
justified in the strong anchoring regime (fixed orientation at
the boundaries) for systems with large line tensions (inclusions
with fixed shape) since their contribution to the free energy is constant.
The one elastic constant approximation (\ref{eq:model}) is not strictly
necessary, since the numerical method to be described below can be applied
to the Frank free energy (\ref{eq:Frank});
the choice of (\ref{eq:model}) however, allows a direct comparison with
the analytical results of \cite{pettey} and provides a stringent test of
our numerics. 

The geometry used in the numerical calculations is shown schematically
in figure~1a.
We consider two circular disks of radius $a$ separated by a distance $D$.
Close to each disk we pin a defect at distances $r_{d_1}$ and $r_{d_2}$
measured from the center of each disk, respectively.  
The system has mirror symmetry with respect to $x$-axis.
Thus, the center of the disks and the accompanying defects are located 
on the $x$-axis.
The smaller circles around the defects are the defect cores (regions
where the nematic order is destroyed \cite{Chaikin}). They are used
as cut-offs for the integral of the elastic free energy density
(\ref{eq:model}).
Lengths are measured in units of $a$ and energy is measured in units of
$K$.

\begin{figure}[ht]
\par\columnwidth=20.5pc
\hsize\columnwidth\global\linewidth\columnwidth   
\displaywidth\columnwidth
\centerline{\epsfxsize=250pt\epsfbox{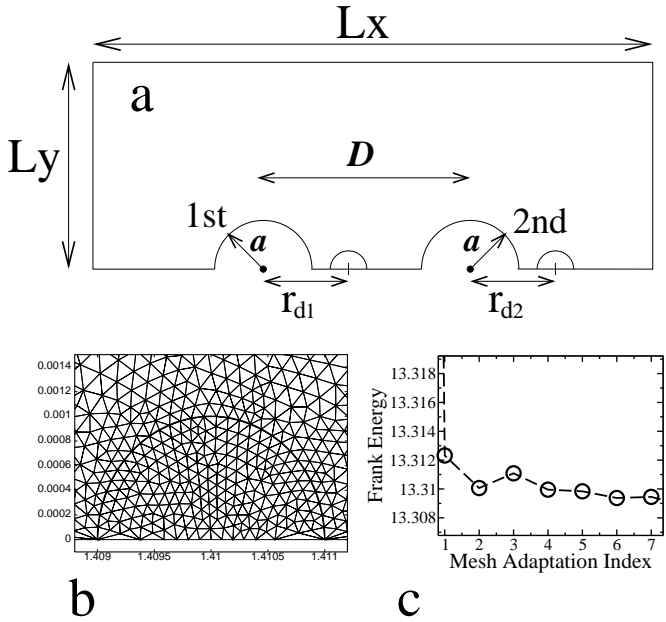}}
\caption
{{\bf a}~: Schematic representation of the geometry with mirror 
symmetry about the $x$-axis. {\bf b}~: Detail of the final
adapted mesh, close to the defect core indicated by the solid
line. The triangulation carefully respects the circular contour.
{\bf c}~: The 2D elastic free energy as a function of the adaptive mesh
iteration index.
After 4 or 5 iterations, the relative error in the computed free energy is
$10^{-4}$.
Figures {\bf b} and {\bf c} correspond to one isolated disk in a
smectic C layer.
Lengths are measured in units of the disk radius $a$ and energy is
measured in units of $K$.}
\label{fig:f1}
\end{figure}

We fixed the nematic orientation, $\theta$, at the physical boundaries.
The far field director was taken parallel to the $x$ axis
($\theta=0$) and homeotropic boundary conditions were imposed at the disk
perimeters.
The solution for the director field is found by minimizing the total
free energy of the system, using standard numerical procedures
\cite{press}.
The major difficulty in the numerical problem stems from the large
difference in the length scales set by the disk and the defect. It was
overcome by using adaptive meshing techniques. With the help
of a 2D mesh generator \cite{george} ({\sl BL2D} package), a first
triangulation
respecting the predefined physical boundaries, as well as the defect
cores, is constructed. 
A final grid for the region close to a defect is shown in figure~1b and
was obtained after several adaptation loops. Note that in addition to the
outer boundaries (figure~1a), the triangulation follows the perimeter of 
the defect core. 
Using standard numerical procedures the elastic free energy is minimized
in a region that includes the defect core. The elastic free energy of
the core region is subtracted at the end of the calculation. The core 
energy was set to zero \cite{Chaikin}.

Adapted meshes may be generated if the Hessian of the solution,
$\partial^2\theta/\partial x\partial y$, is known.
Far from the disks, $\theta$ varies slowly and the triangles
can be large. By contrast, close to the defects, the Hessian is large
(it diverges at $r_d$) and the triangles have to be made several orders of
magnitude smaller.
The Hessian, ${\cal H}_{ij}^k$, at vertex $k$ is estimated \cite{hessian}
by using the following 'weak' definition (\cite{george}, p. 349) 
\begin{equation}
{\cal H}_{ij}^k=\frac{-\int \frac{\partial\theta}
{\partial x_i}\frac{\partial v^k}{\partial x_j}dS}{\int v^k dS}.
\end{equation}
where the integral is over the region occupied by the smectic layer.
$v^k$ is the piecewise linear hat function associated with
vertex $k$ ($v^k=1$ at vertex $k$ and $v^k=0$ elsewhere),
$x_1=x$ and $x_2=y$.
To construct an anisotropic non-uniform mesh,
we require a metric map that distributes the interpolation error,
i.e., the difference between the interpolated function and the exact
solution, in a uniform fashion \cite{george}.
The required map, ${\cal M}_{ij}^k$, at vertex $k$ is a two
dimensional positive definite matrix given by
\begin{equation}
{\cal M}^k=c_0{\cal O}\left(\begin{array}{cc}
|\lambda_1|&0\\
0&|\lambda_2|
\end{array}\right){\cal O}^{-1}
\end{equation}
where ${\cal O}$ is the orthogonal matrix that diagonalizes
the Hessian ${\cal H}^k$, and $\lambda_1$ and $\lambda_2$
the corresponding eigenvalues. The constant $c_0$
controls the number of mesh triangles.
To avoid problems caused by the divergence of the Hessian, the eigenvalues
are bounded by
\begin{equation}
\lambda_i=\min(\max(|\lambda_i|,\frac{1}{c_0h^2_{max}}),\frac{1}{c_0h^2_{min}})
\end{equation}
where $h^2_{min}$ and $h^2_{max}$ are the 
minimal and maximal mesh edge lengths.

As a test of the numerics, we considered one single disk in a
smectic-$C$ film. We calculated the free energy as a function of the
defect distance $r_d$, for a dipole parallel to the far field
director. The equilibrium free energy  
was found for a defect at $r_d=1.41\pm 0.01$, in good agreement with the
analytical result for small defects ($r_d=\sqrt 2 $) \cite{pettey} and in
line with the experimental result for inclusions of cholesteric droplets
in smectic-$C^*$ films ($r_d=1.4 \pm 0.1$) \cite{cluzeau}.

\section{Results}

In the calculations, we used a rectangular integration region of size
$L_x\times L_y=50\times 20$ (see figure~1a).
With the adaptive mesh technique we have studied defects with cores as
small as $10^{-3}$, with a minimal mesh edge length of $10^{-4}$.
During the adaptation loops this limit was reached close to and inside the
cores, providing a good discretization of the core region (see
figure~1b). The final mesh configuration contained up to $10^4$ vertices.
After 4 or 5 iterations of the adaptive mesh procedure, the relative error
of the free energy is of the order of $10^{-4}$ (see figure~1c).
For a particular geometry, the central part of the final mesh is shown in
figure~2a.
In figure~2b, a gray scale was used to plot the corresponding equilibrium
tilt angle, $\theta(x,y)$. White is $\theta=0$ and black $\theta=\pi$.

\begin{figure}[ht]
\par\columnwidth=20.5pc
\hsize\columnwidth\global\linewidth\columnwidth   
\displaywidth\columnwidth
\centerline{\epsfxsize=250pt\epsfbox{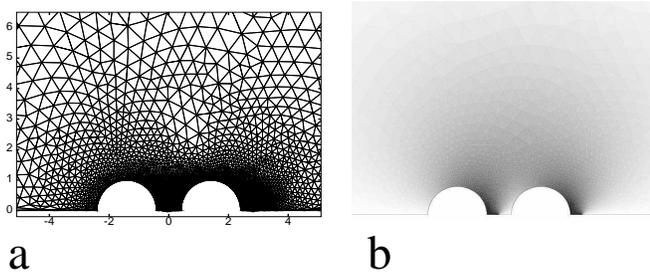}}
\caption
{{\bf a}~: Central part of the final adapted mesh, for a disk separation
corresponding to the minimum free energy ($L_x\times L_y=50\times 20$).
{\bf b}~: In the same region, the equilibrium tilt angle, $\theta(x,y)$, 
is plotted using a gray scale.
White is $\theta=0$ and black $\theta=\pi$.
Lengths are in units of the particle radius $a$.}
\label{fig:f2}
\end{figure}

Near the core (in a small region of size $<10^{-2}$)
the elastic distortion field is determined by the defect alone.
We have checked numerically that a change in the core size
from $\xi_1$ to $\xi_2$ 
leads to a free energy difference of $\pi K \log(\xi_1/\xi_2)$
\cite{pettey}, independent of the position of the disks.

For these systems (small defect cores) we found that the separation of the
disks had no influence 
on the position of the outer defect, $r_{d_2}$~: the minimum free energy
was obtained for an outer defect at $r_d^0$, the position of the defect
for an isolated disk, i.e., independent of the presence of the first disk.
In what follows we set $r_{d_2}=1.41$.

The elastic free energy as a function of the position of the
inner defect, $r_{d_1}$, is shown in figure~3a, for several disk
separations. The lowest free energy was found for a disk separation $D=2.82$, 
with an inner defect at $r_{d_1}=1.41$.

\begin{figure}[ht]
\par\columnwidth=20.5pc
\hsize\columnwidth\global\linewidth\columnwidth   
\displaywidth\columnwidth
\centerline{\epsfxsize=250pt\epsfbox{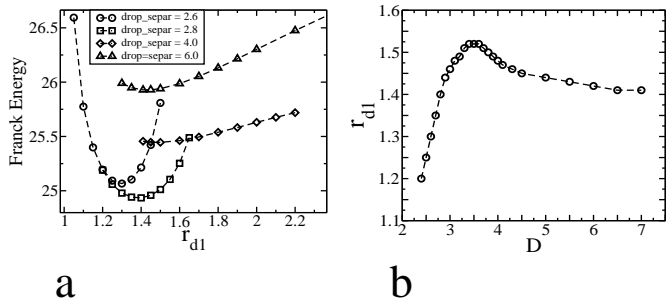}}
\caption
{{\bf a}~: The 2D elastic free energy as a function of the 
position of the inner defect $r_{d_1}$, for several disk
separations. The lowest free energy corresponds
to a disk separation $D=2.82$ and an inner defect at 
$r_{d_1}=1.41$.
{\bf b}~: The position of the inner defect, $r_{d_1}$,
that minimizes the elastic free energy as a function of the disk 
separation, $D$.
Lengths are in units of the disk radius $a$ and energy in
units of $K$.}
\label{fig:f3}
\end{figure}

In figure~3b we plot the position of the inner defect, $r_{d_1}$,
that minimizes the elastic free energy, as a function of the disk
separation, $D$. We identify three regimes.
For small disk separations ($2<D\leq2.82$) the position of the
inner defect, $r_{d_1}$, is symmetrical with respect to the disks and scales 
linearly with $D$, namely $r_{d_1}=D/2$, i.e., the 
defect is at the mid-point between the disks.
In the intermediate regime ($2.82<D<6.5$) the defect position varies
non-monotonically with D, reaching a maximum, $r_{d_1}=1.52$, at
$D=3.5$.
Finally, for large disk separations ($D>6.5$) the position of the
inner defect is independent of D and is given by that of an isolated
disk, $r_{d_1}=1.41$, i.e., it is independent of the presence of the 
second disk.

Note that the minimal free energy corresponds to a symmetrical
configuration: the inner defect is at a distance from the first disk
identical to the distance of the outer defect from the second disk. 

Finally, we plot the elastic free energy as a function of the disk
separation, $D$, in figure~4a. 
The free energy exhibits a pronounced
minimum at $D=2.82\pm 0.01$, close to the value $2.6\pm 0.2$
observed experimentally by Cluzeau et al. \cite{cluzeau}, for chains
of (polydispersed) circular droplets. Note that the experimental estimate
of D, if taken without error bars, suggests that the inner defect is
nucleated at a distance smaller than the distance corresponding to
isolated droplet-defect pairs. This configuration, however, is
ruled out by our model (see figure~3a, open circles and squares). 
Within the (large) error bars we conclude that the experimental 
result is in line with the theoretical prediction: the lowest free
energy corresponds to a symmetrical configuration, with the inner defect
at a distance from the disks that is identical to the distance of
deffects in isolated disk-defect pairs. 

Admitedly, some effects were left out from our calculation, the most
important of which is arguably that of fluctuations. In fact, fluctuations  
were predicted to be important in 2D \cite{pettey} but were reported to be 
negligible in the rather thick films studied by Cluzeau et
al. \cite{cluzeau}. 
It is possible that many-body interactions may account for a reduction 
in disk separation for disks in chains, when compared to the
equilibrium separation of a pair of disks. This could be checked within 
our model by calculating the free energy of three interacting disks. 
The experimental results, however, suggest that this effect, if present, is 
of the order of the experimental error.
A smaller disk separation, than that predicted by our model, may
also result from surface contributions that were 
oversimplified in (\ref{eq:model}). In fact, deviations from strong
homeotropic anchoring at the disks reduce the effective size of the
inclusions, leading to a smaller equilibrium distance between disks
\cite{galerne}.

The signature of the dipolar interaction for large disk separations
may be seen in figure~4b.
Here, the circles are the (calculated) force between disks, as a function 
of their separation. The force is clearly a power law that decays as $D^{-3}$.
Note that the numerical results for the force are less accurate than
those of the free energy, since they were obtained by numerical
differentiation of the latter, figure~4a.

\begin{figure}[ht]
\par\columnwidth=20.5pc
\hsize\columnwidth\global\linewidth\columnwidth   
\displaywidth\columnwidth
\centerline{\epsfxsize=250pt\epsfbox{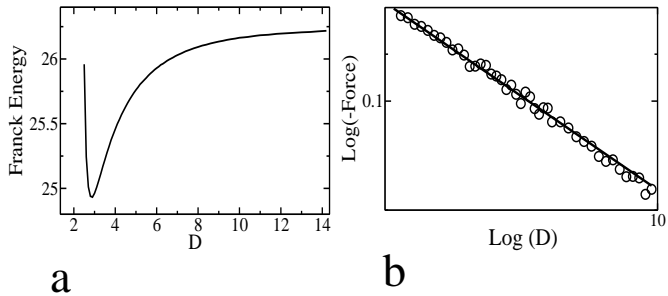}}
\caption
{{\bf a}~: The 2D elastic free energy as a function of the disk
separation $D$.
The free energy has a pronounced minimum at $D=2.82\pm 0.01$.
{\bf b}~: Log-log plot of the effective attractive force between disks
for large separations $D$ ($D>5$). 
Circles are the numerical values obtained by differentiation of
the free energy shown in {\bf a}.
The line is $D^{-3}$.
Lengths are in units of the disk radius $a$ and energy is in
units of $K$.}
\label{fig:f4}
\end{figure}

\section{Summary and Conclusion}

We studied the interaction of two circular disks immersed in smectic-$C$
films.
For strong homeotropic anchoring the circular disks nucleate topological
defects of the 2D director field.
We implemented a finite elements, with adaptive meshing, numerical method 
capable of dealing with the variation of the director field over the
widely different length scales set by the disks and the defects.
The method allows the construction of well defined core regions around the
defects and leads to very accurate results for the elastic free energy.

We have verified the long-range dipolar character of the interaction 
between disk-defect pairs, predicted by Pettey et al. \cite{pettey}.
In addition, we have calculated the interaction at small disk separations,
and found that at static equilibrium, the distance between the centers of 
the disks is equal to $D=2.82\pm 0.01$.

Finally, we have determined the position of the inner defect, as a
function of the separation between disks.
For small disk separations the defect is located symmetrically at the
mid-point between the disks. 
At intermediate separations, this symmetry is broken and the position of
the defect reaches a maximum $r_{d_1}=1.52$ for a disk separation, 
$D=3.5$. 
Finally, at large disk separations the defect relaxes slowly to the
equilibrium position, $r_d^0$, of an isolated disk.

We intend to use the numerical method described in this paper to study
the properties of other reverse nematic emulsions in 2 and 3D.
In particular, we will consider the quadrupolar defects, recently studied
by Fukuda and Yokoyama \cite{fukuda} using a tensor order parameter
formalism. Although these authors concluded that the dipolar configuration 
is dynamically unstable, and the quadrupolar configuration is the only one 
allowed in 2D nematics, the energetics of these two types of pinned
defects has not been compared.

{\bf Acknowledgments}

PP thanks J. M. Tavares for get him 
interested in this problem and for helpful discussions.
We also thank Francisco de los Santos for useful comments.

We acknowledge the support of the Funda\c c\~ao para a 
Ci\^encia e a Tecnologia (FCT) through a running grant (Programa
Plurianual) and  grants No. PRAXIS XXI/BPD/16309/98 (PP) and 
No. SFRH/BPD/1599/2000 (M.T.).

\end{multicols}


\begin{thebibliography}{99}
\bibitem{hessian}
Since we use linear finite elements interpolation -- $\theta$ is given at 
the vertices of a triangle and linearly interpolated inside -- the
lowest order discrete Hessian is zero.
\end{thebibliography}
\end{document}